\documentclass{pasj00}

\newcommand{\swift}{\textit {Swift} }

\begin{document}
\SetRunningHead{K. Inayoshi \& R. Tsutsui}{}
\Received{2011 February 13}
\Accepted{2011 March 29}

\title{Testing two-component jet models of GRBs with orphan afterglows  
}

\author{Kohei \textsc{Inayoshi}, Ryo \textsc{Tsutsui}}
\affil{Department of Physics, Kyoto University, Kyoto 606-8502, Japan}
\email{inayoshi@tap.scphys.kyoto-u.ac.jp, tsutsui@tap.scphys.kyoto-u.ac.jp}


\maketitle

\begin{abstract}
In the \swift era, two-component jet models were introduced to explain the complex temporal profiles 
and the diversity of early afterglows. 
In this paper, we concentrate on the two-component jet model; 
first component is the conventional afterglow and 
second is the emission due to the late internal dissipation such as the late-prompt emission. 
We suggest herein that the two-component jet model can be probed 
by the existence of two optical peaks for orphan GRB afterglows.
Each peak is caused by its respective jet as its relativistic beaming cone widens to encompass the off-axis line of sight.
Typically, the first peak appears at $10^4-10^5$ s and the second at $10^5-10^6$ s. 
Furthermore, we expect to observe a single, bright X-ray peak at the same time as the first optical peak.
Because orphan afterglows do not have prompt emission, it is necessary to monitor the entire sky every $10^4$ s in the X-ray regime. 
We can test the model with orphan afterglows through the X-ray all-sky survey collaboration and 
by using ground-based optical telescopes.
\end{abstract}

\section{Introduction}
The complex temporal profiles of early X-ray afterglows discovered by 
the \swift on-board X-ray telescope (XRT) were unexpected in the pre-\swift era. 
Explaining these results is one of the most difficult problems in the field of gamma-ray bursts (GRBs) (Evans et al. 2009). 
Most X-ray afterglows have the following three phases: steep decay, plateau, and normal decay. 
During the steep-decay phase, the X-ray afterglow decays with a slope of approximately $\sim -3$ 
or steeper: this phase extends up to $10^2-10^3$ s.
The most popular interpretation of this phase attributes it to the tail of 
the prompt emission (Kumar \& Panaitescu 2000; Zhang et al. 2006;
Yamazaki et al. 2006). 
The plateau phase follows the steep-decay phase and begins with 
a  slope of approximately $\sim -0.5$ or flatter and covers the time range $10^3-10^4$ s, 
during which a temporal break occurs. 
The normal decay phase then begins; the afterglow breaks 
and decays with a slope of approximately $\sim -1.2$, which is in agreement with 
the predictions of the afterglow model (M{\'e}sz{\'a}ros \& Rees 1997; 
Sari et al. 1998). 

However, two major problems exist with this interpretation of X-ray afterglows. 
The first is that the plateau phase is not completely explained by the external-shock models.
To address this discrepancies, various models have been proposed, such as the energy-injection model 
(Nousek et al. 2006; Zhang et al. 2006; Granot \& Kumar 2006), 
the inhomogeneous or two-component jet model (Toma et al. 2006; Eichler \& Granot 2006; Granot et al. 2006), 
the time-dependent microphysics model (Ioka et al. 2006; Granot et al. 2006; Fan \& Piran 2006), 
and the prior activity model (Ioka et al. 2006; Yamazaki 2009). 
The second problem is that the conventional afterglow models (M{\'e}sz{\'a}ros \& Rees 1997; 
Sari et al. 1998) cannot explain the difference between the epoch of the breaks in light curves in the 
X-ray and optical observations. 

Ghisellini et al. (2009) reported that the observed light curves are well explained 
by a two-component (early and late) jet model.
Early component is responsible for the conventional afterglow emission and late component for the plateau phase.
When the latter dominates the former, the X-ray light curve exhibits a plateau phase. 
However, there is little evidence that the early and late jet models are correct 
and no way to distinguish these models from other models such as the energy-injection model 
or the time-dependent micro-physics model. 
We show that observations of orphan afterglows are crucial for testing a part of the early and late jet models 
such as the late-prompt emission model (Ghisellini et al. 2007).

In this study, we calculate the light curves of orphan afterglows for the late-prompt emission model.
We find that the two components can be distinguished in the optical regime by observing from off-axis. 
Therefore, if the orphan afterglows present two optical peaks, this fact becomes a strong evidence that GRB afterglows 
consist of two components. 
Moreover, we estimate the detection rate of the orphan afterglows on the basis of the late-prompt emission model 
and find that it would be possible to detect them and thereby test this model.

This article is organized as follows: In Section 2, we explain the motivation for considering the two-component (early and late) 
jet models by reviewing the results of Ghisellini et al. (2009) and introduce the late-prompt emission model.
In Section 3, we describe the numerical method to calculate the off-axis light curves.
The results and their qualitative interpretations are presented in Section 4. 
Finally, in Section 5, we summarize and discuss the possibility of observing orphan afterglows.

\section{Two-component jet Model}
We begin with the two-component (early and late) jet model, which explains the complex light curves, especially 
the plateau phase and the chromatic jet break.
The model assumes that the observed GRB light curves are the sum of two components.
The first component is the {\it early emission} from the forward shock 
due to the interaction of a fireball with the inter-stellar medium (conventional afterglow emission). 
The second component is caused by the late internal dissipation which produces the plateau phase. 
With this model, the difference in the break epoch between the X-ray and 
the optical regime is explained by the fact that the origin of the optical flux 
is different from that of the X-ray flux.

Ghisellini et al. (2009) showed that their model explains well the observed light curves; that is, 
the early emission and the late emission due to a central engine activity that lives long after the prompt emission. 
They assumed that the plateau phase is realized by the late emission, which we call the {\it late-prompt emission} which 
can be expressed as the broken power law 
\begin{eqnarray}
L\propto \left\{ \begin{array}{ll}
T^{-\alpha _{1}} &  (T\leq T_a), \\
T^{-\alpha _{2}} &  (T\geq T_a), \\
\end{array} \right.
\end{eqnarray}
where $T_a$ is the time at which the plateau phase ends, and the constants 
are typically $\alpha _1\sim 0$ and $\alpha _2\sim 1.6$ (Ghisellini et al. 2009).

They obtained several parameters for this model by fitting it to 33 GRBs with known redshifts.
As an example, we present one of the 33 fit results in Figure 1. 
The lines show the fit by Ghisellini's model; the red, blue, and black lines represent the early emission, 
the late-prompt emission, and their sum, respectively. 
The solid and dashed lines refer to the X-ray light curves and the optical light curves, respectively. 
The dynamics of this GRB exhibit a plateau phase in the X-ray regime but not in the optical regime.
The sum of these two components describes well the light curve of the plateau phase 
and the normal phase in both bands. 
In this article, we concentrate on the late-prompt emission model and 
suggest a method to verify whether these two components exist.

\section{Numerical Method}
In this section, we describe our calculation of the off-axis light curves of the early emission 
and the late-prompt emission.
Because of the strong beaming effect, the observed off-axis emission does not exhibit 
prompt emission and therein is often called the "orphan" afterglow.
Here, we define the time $T$ as the observed time elapse since the GRB trigger 
if the prompt emission were isotropic.

\begin{figure}[t]
\begin{center}
\rotatebox{0}{\includegraphics[height=65mm,width=80mm]{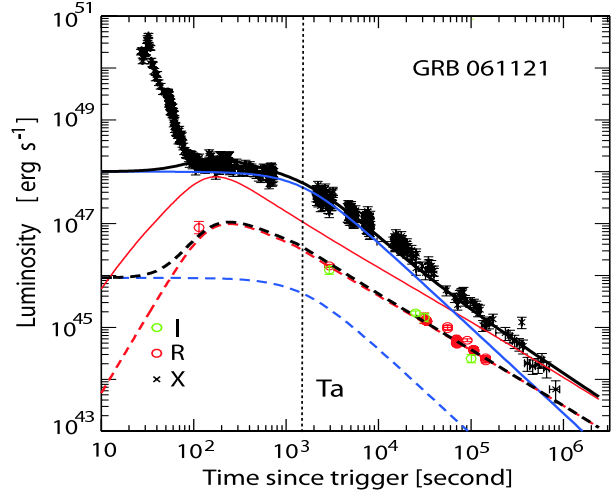}}
\end{center}
\caption{X-ray ($0.3-10$ keV) (black crosses) and optical (green and red circles) light curves. 
The lines show the fit by the two-component jet model (Ghisellini et al. 2009); 
the red, blue, and black lines represent the early emission, 
the late-prompt emission, and their sum, respectively. 
The solid and dashed lines refer to the X-ray light curves and the optical ones, respectively. 
The vertical line corresponds to $T_a$.
The data are taken from Page et al. 2007 (and references therein).}
\label{}
\end{figure}

\subsection{Orphan early emission}
The dynamics of the orphan early emissions have already been predicted by several authors 
(Totani \& Panaitescu 2002; Granot et al. 2002; Granot 2005 ). 
Assume that a photon emitted at lab-frame time $t$ (the frame of the central engine), radius $R$, and angle $\theta$ 
from the line-of-sight reaches the observer at $T$. 
In this case, $T$ is given by $\frac{T}{1+z}=t-\frac{R\cos \theta }{c}$,
where $z$ is the GRB redshift and $R$ is given by
\begin{equation}
R=c\int^{t}_0\beta (t')dt'\simeq ct-\int^R_0\frac{dR'}{2\gamma ^2},
\end{equation}
for $\gamma \gg 1$. 
For simplicity, we assume the "thin shell" case, for which the width and expansion of the shell are neglected in calculating its deceleration. 
The dynamics of the forward shock are determined by the self-similar solution (Blandford \& McKee 1976). 
The Lorentz factor as a function of $R$ is $\gamma (R)\simeq \eta $ for $R<R_{\rm{dec}}$ 
and $\gamma (R)\simeq \eta( R/R_{\rm{dec}})^{-3/2}$ for $R>R_{\rm{dec}}$, 
where $\eta $ is the initial Lorentz factor of the shell (typically $\eta \sim 10^2$) 
and $R_{\rm{dec}}$ is the deceleration radius. 
With these relationships, the flux density in the thin-shell case is given by
\begin{eqnarray}
&&F_{\nu}(T)=\frac{(1+z)}{d_L^2(z)}\nonumber \\
&&\hspace{5mm}\times \int d^4x\delta \Big( 
 t-\frac{T}{1+z}-\frac{R\cos \theta}{c}\Big) \frac{j'_{\nu '}}{\gamma 
 ^2(1-\beta \cos \theta )^2},
\end{eqnarray}
where $d_L(z)$ is the luminosity distance and $L'_{\nu '}$ and $\nu '$ are 
the spectral luminosity in units of erg s$^{-1}$ Hz$^{-1}$ and the frequency in the comoving frame, respectively. 
The quantity $L' _{\nu '}$ may be expressed as a function of $R$ and $\nu '$ (Sari 1998, Granot et al 2002).

By integrating the right-hand side of Eq.(3), we can express the flux 
density as
\begin{equation}
F_{\nu }(T)=\frac{(1+z)}{8\pi d_L^2(z)}\int d\cos \theta 
 \frac{L'_{\nu '}(R)}{\gamma ^3(1-\beta \cos \theta )^3}\frac{\Delta \phi 
 }{2\pi },
\end{equation}
where $\frac{\Delta \phi }{2\pi }$ is the correction factor when the viewing angle 
is larger than the jet opening angle ($\theta _v>\theta _j$). 
Certainly, for $\theta _v=0$, $\frac{\Delta \phi }{2\pi }=1$ 
for $0<\theta <\theta _j$. Outside the jet angle, the correction factor is given by
\begin{eqnarray}
\frac{\Delta \phi }{2\pi }\simeq \left\{ \begin{array}{ll}
0 &  (\theta <\theta _v-\theta _j), \\
\frac{\phi }{\pi } &  (\theta _v-\theta _j<\theta <\theta _v+\theta _j), \\
0 &  (\theta _v+\theta _j<\theta ), \\
\end{array} \right.
\end{eqnarray}
\begin{equation}
\cos \phi =\frac{\cos \theta _j-\cos \theta _v\cos \theta }{\sin \theta 
 _v\sin \theta },
\end{equation}
(Woods \& Loeb 1999, Yamazaki et al. 2003). 
Using these relationships, we can integrate to obtain the flux density of the early emission 
received by the observer at $T$. 

\subsection{Orphan late-prompt emission}
We apply the above formula to the off-axis late-prompt emission. 
We begin with a brief description of the late-prompt emission model (Ghisellini et al. 2007). 
This model assumes that the central engine continues after the prompt emission to create relativistic shells 
with smaller Lorentz factors $\gamma _L(T)$ and lower powers. 
Initially, $\gamma _L\gg \theta _{jL}^{-1}$, where $\theta _{jL}$ is the late-prompt jet opening angle. 
Then, the emission area seen from on-axis observer is the order of $(R/\gamma _L)^2$ 
because of the beaming effect. 
When $\gamma _{L}$ decreases with time and becomes smaller than $\theta _{jL}^{-1}$, the emission area becomes $(R\theta _{jL}) ^2$. 
Thus, the observed luminosity (erg s$^{-1}$) of the late-prompt emission can be expressed as 
\begin{eqnarray}
L\propto \left\{ \begin{array}{ll}
 R^2\Delta R'j'(T)\propto 
 T^{-\alpha _{1}} &  (T\leq T_a), \\
(R\theta _j) ^2\Delta R'\gamma _L^2j'(T)\propto 
 T^{-\alpha _{2}} &  (T\geq T_a), \\
\end{array} \right.
\end{eqnarray}
where $j'(T)$ is the bolometric emissivity in the comoving frame and $T_a$ is the time when $\gamma _L(T_a)=\theta _{jL}^{-1}$. 
When $T=T_a$, one sees the jet break as in the early emission. 
In this model, the time $T_a$ corresponds to the transition from the plateau phase to the normal decay phase. 
Moreover, assuming that $j'(T)$ has a constant slope before and after $T_a$, it is shown that 
$\gamma _L(T)\propto T^{-\Delta \alpha /2}$ ($\Delta \alpha =\alpha _{2}-\alpha _{1}$). 
Note that the time dependence of $\gamma _L$ is phenomenologically introduced to explain the plateau phase. 
In addition, the dissipation processes responsible for the late-prompt emission would be the same as that for 
the prompt emission (e.g., internal shocks). 
Then, this model assumes that the emission occurs at a fixed radius $R$ 
and the observed spectrum is well approximated by the Band function (Band et al. 1993); 
$f(\nu )\propto \nu ^{-\beta _{\rm o}}~(\nu ^{-\beta _{\rm X}})$ for $\nu < \nu_b$ ($\nu > \nu_b$, respectively) 
and the indices are typically $\beta _{\rm o}\sim 0.1$ and $\beta _{\rm X}\sim 1.0$ (Ghisellini et al. 2009).
With these relations, we can obtain the flux density 
\begin{equation}
F_{\nu }(T)=\frac{(1+z)}{8\pi d_L^2(z)}\int d\cos \theta 
 \frac{\tilde f(T, \nu )}{\gamma _L^2(1-\beta _L\cos \theta )^2}\frac{\Delta \phi 
 }{2\pi },
\end{equation}
where $\tilde f(T, \nu )\propto T^{-\alpha _1}f(\nu )$. 
By integrating the right-hand side, we can therefore obtain the off-axis light curves of the late-prompt emission 
just like the orphan early emission. 

\begin{figure}[!t]
\begin{center}
\rotatebox{0}{\includegraphics[height=62mm,width=80mm]{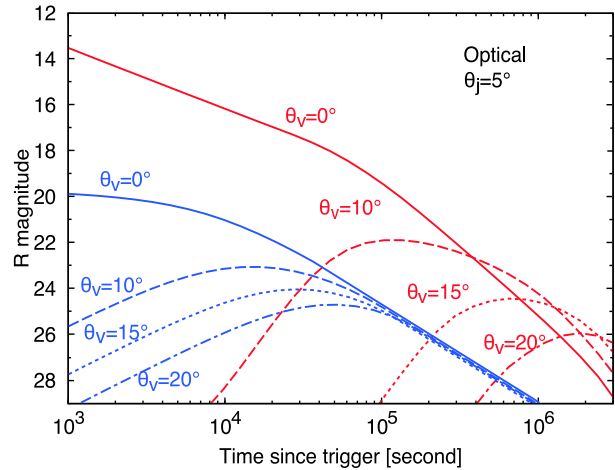}}
\end{center}
\caption{Optical decay curves (R-band) for the early emission (red lines) and 
the late-prompt emission (blue lines) when the early emission dominates the late-prompt emission. 
Two peaks are present, which are distinguished more clearly for larger viewing angles. 
The jet-opening half angle, $\theta _j=\theta _{Lj}=5^\circ $. 
The physical parameters of the jets are $E=10^{52}$ erg, $\epsilon _B=0.01$, $\epsilon _e=0.1$, 
$n=1.0$ cm$^{-3}$, $p=2.4$, $\eta =200$, $z=1$, $\alpha _1=0.1$, $\alpha _2=1.6$, $\beta _{\rm X}=0.9$, $\beta _{\rm o}=0$ 
and $\nu _b=5\times 10^{16}$ Hz, $T_a=10^4$ s, $L_X|_{T_a}=10^{47}$ erg s$^{-1}$. 
These parameters of the jets are taken to be typical values shown in Ghisellini et al. (2009). 
}
\label{}
\end{figure}

\section{Results}
We begin with a short description for the qualitative nature of the off-axis light curves. 
First, we focus on the radiation from the early emission component. 
Because the Lorentz factor of shells producing emission initially is relativistic and the beaming effect is strong, 
the off-axis observer cannot see the emission. 
When the Lorentz factor decreases to $\theta _v^{-1}$, its relativistic beaming cone widens to encompass the off-axis observer.
Then, the radiation can reach the off-axis observer and he observes the emission as a single peak.  
After that, the off-axis light curves undergo a power-law decay like the on-axis dynamics.
For the early emission, the relation between the jet break time $T_j$ and the peak time $T_p$ ($\gamma \simeq \theta _v^{-1}$) 
is given by Totani \& Panaitescu (2002); 
\begin{equation}
T_p=\Big( 5+2\ln \frac{\theta _v}{\theta _j}\Big) \Big( \frac{\theta _v}{\theta _j}\Big) ^2T_j. 
\label{tp}
\end{equation}
When $\theta _v\simeq (2-5)\times \theta _j$, we obtain $T_p\simeq 7(\theta _v/\theta _j)^2T_j$. 

For the late-prompt emission, the relation between the transition time $T_a$ and the peak time $T_{pL}$ ($\gamma _L\simeq \theta _v^{-1}$) 
is given by 
\begin{equation}
T_{pL}=\Big( \frac{\theta _v}{\theta _{jL}}\Big) ^{2/\Delta \alpha }T_a. 
\label{tpL}
\end{equation}
From Eq.(\ref{tp}) and Eq.(\ref{tpL}), we have 
\begin{equation}
\frac{T_{p}}{T_{pL}}=7\Big( \frac{\theta _{jL}}{\theta _{j}}\Big) ^2\frac{T_j}{T_a}, 
\end{equation}
where we set $\Delta \alpha =1$. 
Unfortunately, the magnitude relation of the two jet opening angles 
($\theta _{jL}$ and $\theta _j$) is an uncertainty. 
When the jet opening angle of the late-prompt emission is larger than 
that of the early emission $\theta _{jL}\geq \theta _j$, we obtain $(T_p/T_{pL})>$ a few $(T_j/T_a)$.
Thus, we find $T_p\simeq (10-100)\times T_{pL}$ because typically 
$T_j\simeq 10^5-10^6$ s and $T_a\simeq 10^4$ s from the observational facts. 
If the peak amplitudes of the early emission and the late-prompt emission are the same order, therefore, 
we are able to observe two peaks in the light curve by viewing from off-axis. 
On the other hand, when the late-prompt emission is encompassed by the early emission, 
i.e., $\theta _{jL}<\theta _j$, the peak times of each emission can become the same order ($T_p\sim T_{pL}$) . 
In this case, even if the peak fluxes of two emissions are the same order, 
the two peaks overlap and are difficult to distinguish clearly. 
In this paper, we concentrate on the case of $\theta _{jL}\simeq \theta _j$.

Next, we present the results of our calculation and explain the behavior of the early emission 
and the late-prompt emission. 
Figure 2 and 3 show the optical and X-ray light curves, respectively, for various viewing angles 
($\theta _v=0^\circ$, $10^\circ$, $15^\circ$, and $20^\circ $) relative to the center of the jet ($\theta _j=\theta _{jL}=5^\circ $). 
In Figure 4, we present the observed optical light curves: that is, the sum of the early emission and the late-prompt emission.

\begin{figure}[t]
\begin{center}
\rotatebox{0}{\includegraphics[height=62mm,width=80mm]{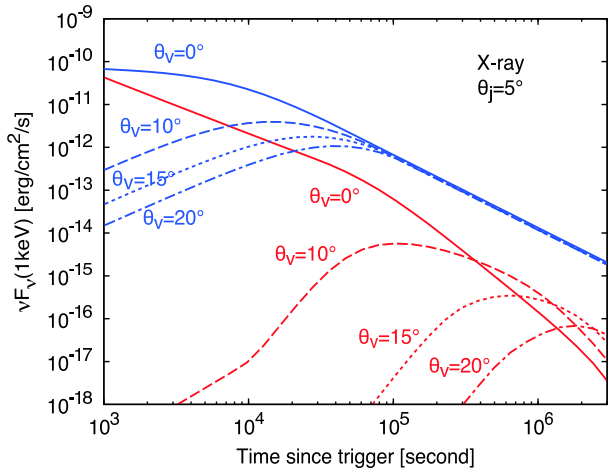}}
\end{center}
\caption{X-ray light curves (1 keV) for the early emission (red lines) and 
the late-prompt emission (blue lines) when the late-prompt emission dominates the early emission. 
The light curves observed off-axis (the early emission plus the late-prompt emission) have a single peak. 
The physical parameters used for the jets are the same as for Figure 2. 
}
\label{}
\end{figure}

In Figure 2, we show the optical light curves for various viewing angle when the early emission (red lines) dominates the 
late-prompt emission (blue lines). 
In this case, as shown in Figure 2, we can observe the optical flux with two peaks due to 
the orphan early emission and the orphan late-prompt emission viewing from off-axis 
because both peak amplitudes are the same order. 
In addition, we find that the two peaks are more clearly distinguished and each peak becomes dimmer for larger viewing angle (see Figure 4). 
By observing two peaks in the optical, therefore, we can verify that GRB light curves consist of two components or not. 

In Figure 3, we show the light curves when the X-ray flux is dominated by the late-prompt emission. 
According to Ghisellini et al. (2009), the X-ray flux is dominated by the late-prompt emission 
or by a mixture of both emissions. 
In this case, the observed off-axis light curves (the sum of red and blue lines for each viewing angle $\theta _v$) 
have a single peak because the early emission is dominated by the late-prompt emission. 
As show in Figure 3, the amplitude ratio (on/off) for the early emission is larger than the amplitude ratio for the late-prompt emission and 
the early emission's peak decreases at a great rate than the late-prompt emission's peak as the viewing angle increases. 
Thus, for larger viewing angle, the late-prompt emission always dominates the early emission and the characteristic that the observed light curves 
have a single peak does not change. 

Furthermore, since the transition at $T_a$ corresponds to a phenomenon like the jet break in this model, 
the peak amplitude observed off-axis is three to five orders of magnitude larger than 
that expected from the conventional models (due to the orphan early emission). 
Thus, we can observe more orphan afterglows than expected from the conventional models if the plateau phase is 
attributed to the late-prompt emission. 
For these reasons, X-ray observation facilitates the detection of orphan GRB afterglows. 

\begin{figure}[t]
\begin{center}
\rotatebox{0}{\includegraphics[height=62mm,width=80mm]{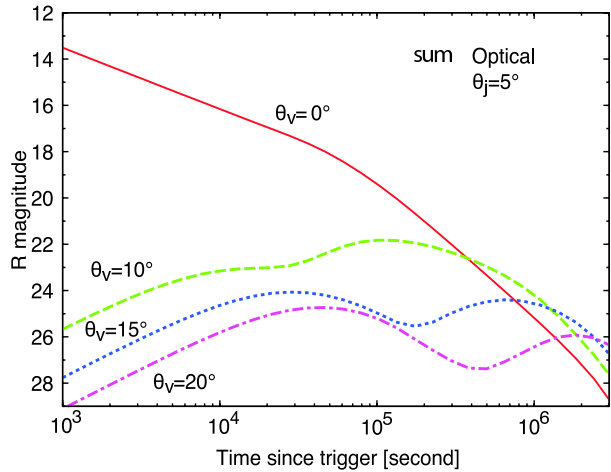}}
\end{center}
\caption{Observed optical light curves (R-band); sum of the early emission and the late-prompt emission.
The physical parameters are the same as in Figure 2.}
\label{}
\end{figure}

\section{Summary \& Discussion}
In this study, we consider the late-prompt emission model and calculate the off-axis observed light curves. 
Then, we find that the off-axis observer is able to see the emission with two peaks in the optical (Figure 4) and 
with a single peak in the X-ray (Figure 3). 
We here suggest a method to test whether two components actually exist or not by applying the characteristics of off-axis light curves. 

In this section, we discuss the possibility of observing the emission from off-axis. 
As seen in Section 4, in the late-prompt model, the single peak in the X-ray regime is roughly as bright as the flux amplitude
at the transition time $T_a$. 
Then, we expect that the emission with a single peak in the X-ray is observable. 
We can easily estimate the detection rate of the X-ray flux viewed off-axis for the late-prompt model 
by using the same samples as in Ghisellini et al. (2009). 
Assuming that the properties of GRBs are not affected by the accuracy with which the redshift is determined, 
we can extrapolate the flux distribution of all \swift bursts from the above samples.
Here, we suppose that the average jet angle of the GRBs is 5$^\circ $. 
Next, the real number $N_{\rm{real}}(F)$ of GRBs per year for flux $F$ is not the number $N_{\rm{obs}}(F)$ of GRBs 
inferred from observations, but $2N_{\rm{obs}}(F)/(1-\cos \theta _j)$. 
From our calculation, we find the relationship between the viewing angle $\theta _v$ and the peak flux for the late-prompt emission 
to be $F_{\rm{peak}}(\theta _v)\propto \theta _v ^{-2.09}$ ($\theta _v >\theta _j$). 
The detection rate $N_{\rm{exp}}$ is then estimated as 
\begin{equation}
N_{\rm{exp}}=\int \frac{d\phi d\cos \theta _v}{4\pi } N_{\rm{real}}(F_{\rm{peak}}(\theta _v)>F_{\rm{lim}}),
\end{equation}
where $F_{\rm{lim}}$ is the instrument sensitivity. 

Because the GRB observed off-axis do not have the prompt emission, 
the all-sky survey that monitors the whole sky is necessary to 
detect the single peak in the X-ray. 
For example, such peak with duration time of $10^4$ s can be detected by the current X-ray 
survey missions with Monitor of All-sky X-ray Image (MAXI) (Matsuoka et al. 2009).  
The sensitivity of MAXI is 20 mCrab ($7\times 10^{-10}$ erg~cm$^{-2}$~s$^{-1}$ over the energy band  $2-30$ keV)
for observations during a single orbit. 
Because MAXI observes a point of sky every $90$ min, it is suitable for detecting the peak. 
In this case, the detection rate by MAXI is estimated to be $\sim 0.41$ events per year. 
As seen Figure 3, the peak amplitudes of the orphan late-prompt emission are a little lower than the MAXI sensitivity. 
In case of  the orphan late-prompt emission, however, the number of GRBs whose relativistic beaming cones encompass the observer increases 
by a factor of $(1-\cos \theta _v)/(1-\cos \theta _j)$. 
The event rate we estimate is as much as the event rate that MAXI detects on-axis afterglows without prompt emissions.
This forecast is an important outcome of the late-prompt model. 
Because the angular resolution of MAXI is about 0.1 arcminute, we can follow up the afterglows 
with ground-based optical telescopes (e.g., Gamma-Ray Burst Optical/Near-Infrared Detector; GROND) 
and observe emission with two peaks. 
Typical amplitudes of the first peak are about $24$ to $25$ in R-band magnitude. 
Because this sensitivity is of the same order as that of GROND (Greiner et al. 2008), 
it should be possible to observe the first optical peak. 
Although the second peak is slightly dimmer than the first, we consider that after a few days, the second peak 
can be observed by GROND or the larger optical telescopes. 
Seven bands of GROND observation help to distinguish the afterglows from other variables.
The second peak, which has a synchrotron spectrum because of the external shock, can be distinguished from supernovae. 

In this study, we calculate the off-axis light curves in the late-prompt model by integrating the analytical expressions. 
However, in recent study, van Eerten, Zhang \& MacFadyen (2010) show that the off-axis light curves calculated analytically are 
different from the results with a two-dimensional axisymmetric hydrodynamics simulation. 
Then, it is significant to calculate the off-axis late-prompt emission in order to discuss whether two peaks are distinguishable. 
Furthermore, the late-prompt emission model assumes the radius $R$ to be constant. 
If the radius expands of the emission area, the peak time $T_{pL}$ for the late-prompt emission more delays in the same way as 
the peak time $T_p$ for the early emission as the viewing angle increases. 
In this case, it may be, therefore, difficult to distinguish the optical two peaks in the optical with the observation of orphan afterglows. 

Finally, we mention other late-internal dissipation models. 
In these models, the transition time $T_a$ does not necessarily correspond to 
the jet break but to the time scale 
for the matter accretion (Kumar et al. 2008) or 
the spin-down time scale for the magnetar scenarios 
(Zhang \& M{\'e}sz{\'a}ros 2001; Thompson et al. 2004). 
If this is the case, then the Lorentz factor evolution history
of the late outflow is unpredictable, and the predicted 
orphan afterglow light curves would be subject to even larger uncertainties.
In future works, we should discuss other possibilities for 
testing two-component jet models. 
Nevertheless, if we observe a characteristic light curve, 
which has a bright single peak in the X-ray and 
two peaks in the optical, we obtain an evidence that 
GRBs light curves are composed of the early and late jet, 
and may restrict other models than the early and late jet model.

\bigskip
We would like to thank T. Nakamura for his continuous encouragement and 
K. Murase, K. Yagi, M. Hayashida, S. Inoue, Y. Inoue, and Y. Doleman for 
useful discussions. 
This work is supported in part by the Grant-in-Aid
for the global COE program {\it The Next Generation of Physics, Spun from Universality and Emergence} at Kyoto University.
RT is supported by a Grant-in-Aid for the Japan Society for the Promotion of Science (JSPS) Fellows
and is a research fellow of JSPS.


\end{document}